\documentclass{article}
\usepackage{amsmath,graphicx,mlspconf}
\usepackage{subcaption}
\usepackage{amssymb}
\usepackage{amsfonts}
\usepackage{mathtools}
\usepackage{multirow}
\usepackage{bm}
\usepackage{nicefrac}
\usepackage{tabularx}
\usepackage{booktabs}
\usepackage{pifont}
\usepackage{amsthm}
\usepackage{extarrows}
\usepackage[bookmarks=false]{hyperref}
\usepackage{balance}
\usepackage[ruled,vlined]{algorithm2e}
\usepackage{array}

\makeatletter
\def\hlinewd#1{%
  \noalign{\ifnum0=`}\fi\hrule \@height #1 \futurelet
   \reserved@a\@xhline}
\makeatother

\SetKwInput{KwInput}{Input}                
\SetKwInput{KwOutput}{Output}              

\theoremstyle{definition}
\newtheorem{definition}{Definition}[section]

\newcommand{\sudo}{SuDoRM-RF }
\newcommand{\sudodot}{SuDoRM-RF}
\newcommand{\sudol}{SuDoRM-RF 1.0x }
\newcommand{\sudom}{SuDoRM-RF 0.5x }
\newcommand{\sudos}{SuDoRM-RF 0.25x }

\newcommand\srelu{\operatorname{ReLU}}

\newcommand\lp{\left(}
\newcommand\rp{\right)}

\newcommand\y{\mathbf{y}}
\newcommand\x{\mathbf{x}}
\newcommand\T{T}
\newcommand\CE{C_{\mathcal{E}}}
\newcommand\Cout{C}
\newcommand\Cin{C_U}
\newcommand\Kin{K_U}
\newcommand\Sin{S_U}

\newcommand\KE{K_{\mathcal{E}}}

\newcommand\R{\mathbb{R}}
\newcommand\vx{\mathbf{v}_{\mathbf{x}}}
\newcommand\evi{\widehat{\mathbf{v}}_{i}}
\newcommand\emi{\widehat{\mathbf{m}}_{i}}
\newcommand\esi{\widehat{\mathbf{s}}_{i}}

\newcommand\s{\mathbf{s}}

\title{Sudo rm -rf: Efficient Networks for Universal Audio Source Separation}
\name{Efthymios Tzinis${^\natural}$ \quad Zhepei Wang${^\natural}$ \quad Paris Smaragdis$^{\natural,\flat}$\thanks{Supported by NSF grant \#1453104. A Titan V used for this research was donated by the NVIDIA Corporation.}}
\address{$^\natural$ University of Illinois at Urbana-Champaign \\
$^\flat$ Adobe Research\\ }

\begin{document}
\ninept

\maketitle

\begin{abstract}
In this paper, we present an efficient neural network for end-to-end general purpose audio source separation. Specifically, the backbone structure of this convolutional network is the SUccessive DOwnsampling and Resampling of Multi-Resolution Features (\sudodot) as well as their aggregation which is performed through simple one-dimensional convolutions. In this way, we are able to obtain high quality audio source separation with limited number of floating point operations, memory requirements, number of parameters and latency. Our experiments on both speech and environmental sound separation datasets show that \sudo performs comparably and even surpasses various state-of-the-art approaches with significantly higher computational resource requirements.  
\end{abstract}
\section{Introduction}
\label{sec:intro}

The advent of the deep learning era has enabled the effective usage of neural networks towards single-channel source separation with mask-based architectures \cite{huang2014deep}. Recently, end-to-end source separation in time-domain has shown state-of-the-art results in a variety of separation tasks such as: speech separation \cite{luo2019convTasNet, luo2019dual}, universal sound separation \cite{kavalerov2019universal, tzinis2019improving} and music source separation \cite{defossez2019demucs}. The separation module of ConvTasNet \cite{luo2019convTasNet} and its variants \cite{kavalerov2019universal, tzinis2019improving} consist of multiple stacked layers of depth-wise separable convolutions \cite{sifre2014depthwiseseparable} which can aptly incorporate long-term temporal relationships. Building upon the effectiveness of a large temporal receptive field, a dual-path recurrent neural network (DPRNN) \cite{luo2019dual} has shown remarkable performance on speech separation. Demucs \cite{defossez2019demucs} has a refined U-Net structure \cite{ronneberger2015original_unet} and has shown strong performance improvement on music source separation. Specifically, it consists of several convolutional layers in each a downsampling operation is performed in order to extract high dimensional features. A two-step approach has been introduced in \cite{tzinis2019two} and showed that universal sound separation models could be further improved when working directly on the latent space and learning the ideal masks on a separate step.

Despite the dramatic advances in source separation performance, the computational complexity of the aforementioned methods might hinder their extensive usage across multiple devices. Specifically, many of these algorithms are not amenable to, e.g., embedded systems deployment, or other environments where computational resources are constrained.  Additionally, training such systems is also an expensive computational undertaking which can amount to significant costs.

Several studies, mainly in the image domain, have introduced more efficient architectures in order to overcome the growing concern of large models with high computational requirements. Models with depth-wise separable convolutions \cite{sifre2014depthwiseseparable} have shown strong potential for several image-domain tasks \cite{chollet2017xception_depthwiseseparable} while significantly reducing the computational requirements. Thus, several variants such as MobileNets \cite{howard2017mobilenets} have been proposed for deep learning on edge-devices. However, convolutions with a large dilation factor might inject several artifacts and thus, lightweight architectures that combine several dilation factors in each block have been proposed for image tasks \cite{mehta2019espnetv2}. More recent studies propose meta-learning algorithms for optimizing architecture configurations given specific computational resource and accuracy requirements \cite{yu2019slimmablenets, cai2019onceandforall}.

Despite the recent success on low-resource architectures on the image domain, little progress has been made towards proposing efficient architectures for audio tasks and especially source separation. In \cite{kalchbrenner2018efficient_audiosynthesis} a WaveRNN is used for efficient audio synthesis in terms of floating point operations (FLOPs) and latency. Other studies have introduced audio source separation models with reduced number of trainable parameters \cite{luo2019dual, maldonado2020lightweight} and binarized models \cite{kim2018bitwise}. In this study, we propose a novel efficient neural network architecture for audio source separation while following a more holistic approach in terms of computational resources that we take into consideration (FLOPs, latency and total memory requirements). Our proposed model performs SUccessive DOwnsampling and Resampling of Multi-Resolution Features (\sudodot) using depth-wise convolutions. By doing so, \sudo exploits the effectiveness of iterative temporal resampling strategies \cite{haris2018deepbackprojectionSuperResolution} and avoids the need of multiple stacked dilated convolutional layers \cite{luo2019convTasNet}. We report a separation performance comparable or even better to several recent state-of-the-art models on speech and environmental sound separation tasks with significantly lower computational requirements. Our experiments suggest that \sudo models a) could be deployed on devices with limited resources, b) be trained significantly faster and achieve good separation performance and c) scale well when increasing the number of parameters. Our code is available online\footnote{Code: \href{https://github.com/etzinis/sudo\_rm\_rf}{https://github.com/etzinis/sudo\_rm\_rf}}.

\section{Sudo rm -rf network architecture}
\label{sec:net_arch}
\begin{figure}[!htb]
    \centering
      \includegraphics[width=\linewidth]{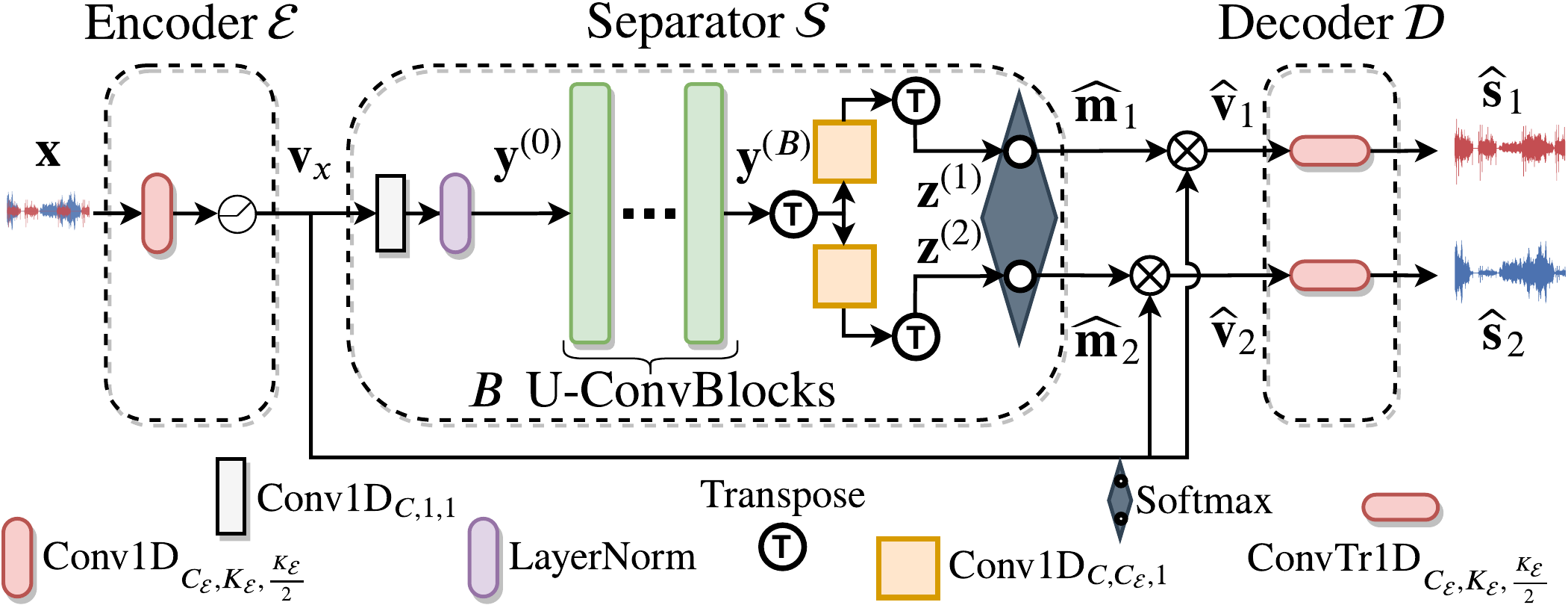}
      \caption{\sudo architecture for separating two sources.}
      \label{fig:sudormrf}
\end{figure}

On par with many state-of-the-art approaches in the literature \cite{luo2019convTasNet, tzinis2019two, luo2019dual, defossez2019demucs}, \sudo performs end-to-end audio source separation using a mask-based architecture with adaptive encoder and decoder basis. The input is the raw signal from a mixture $\textbf{x} \in \R^T$ with $T$ samples in the time-domain. First we feed the input mixture $\textbf{x}$ to an encoder $\mathcal{E}$ in order to obtain a latent representation for the mixture $\vx = \mathcal{E} \left( \textbf{x} \right) \in \R^{\CE \times L}$. Consequently the latent mixture representation is fed through a separation module $\mathcal{S}$ which estimates the corresponding masks $\emi \in \R^{\CE \times L}$ for each one of the $N$ sources $\s_1, \cdots, \s_N \in \R^T$ which constitute in the mixture. The estimated latent representation for each source in the latent space $\evi$ is retrieved by multiplying element-wise an estimated mask $\emi$ with the encoded mixture representation $\vx$. Finally, the reconstruction for each source $\esi$ is obtained by using a decoder $\mathcal{D}$ to transform the latent-space $\evi$ source estimates back into the time-domain $\esi = \mathcal{D} \left( \evi \right)$. An overview of the \sudo architecture is displayed in Figure \ref{fig:sudormrf}. The encoder, separator and decoder modules are described in Sections \ref{sec:net_arch:encoder}, \ref{sec:net_arch:separator} and \ref{sec:net_arch:decoder}, respectively. For simplicity of our notation we will describe the whole architecture assuming that the processed batch size is one. Moreover, we are going to define some useful operators of the various convolutions which are used in \sudodot.

\begin{definition}
$\operatorname{Conv1D}_{C, K, S}:\R^{C_{in} \times L_{in}} \rightarrow \R^{C \times L}$ defines a kernel $\mathbf{W} \in \R^{C \times C_{in} \times K}$ and a bias vector $\mathbf{b} \in \R^{C}$. When applied on a given input $\mathbf{x} \in \R^{C_{in} \times L_{in}}$ it performs a one-dimensional convolution operation with stride equal to $S$ as shown next:
\begin{equation}
\label{eq:conv1d}
    \begin{gathered}
    \operatorname{Conv1D}_{C, K, S}\lp \x \rp_{i,l} = \mathbf{b}_i + \sum_{j=1}^{C_{in}} \sum_{k=1}^{K} \mathbf{W}_{i,j,k}\cdot \x_{j, S\cdot l- k},
    \end{gathered}
\end{equation}
where the indices $i$, $j$, $k$, $l$ denotes the output channel, the input channel, the kernel sample and the temporal index, respectively.  Note that without loss of generality and performing appropriate padding, the last dimension of the output representation would be $L = \left \lfloor{\nicefrac{L_{in}}{S}}\right \rfloor$. 
\end{definition}

\begin{definition}
$\operatorname{ConvTr1D}_{C, K, S}:\R^{C_{in} \times L_{in}} \rightarrow \R^{C \times L}$ defines a one-dimensional transpose convolution. Since any convolution operation could be expressed as a matrix multiplication, transposed convolution can be directly understood as the gradient calculation for a regular convolution w.r.t. its input \cite{Simonyan2013DeepIC_transposeconvolution}.
\end{definition}

\begin{definition}
$\operatorname{DWConv1D}_{C, K, S}:\R^{C_{in} \times L_{in}} \rightarrow \R^{C \times L}$ defines a one-dimensional depth-wise convolution operation \cite{sifre2014depthwiseseparable}. In essence, this operator defines $G=C_{in}$ separate one-dimensional convolutions $\mathcal{F}_i = \left[ \operatorname{Conv1D}_{C_{G}, K, S} \right]_i$ with $i \in \{1, \cdots, G \}$ where $C_{G} = \left \lfloor{\nicefrac{C}{G}}\right \rfloor$. Given an input $\mathbf{x} \in \R^{C_{in} \times L_{in}}$ the $i$th one-dimensional convolution contributes to $C_{G} = \left \lfloor{\nicefrac{C}{G}}\right \rfloor$ output channels by considering as input only the $i$th row of the input as described below: 
\begin{equation}
\label{eq:dwconv1d}
    \begin{gathered}
    \operatorname{DWConv1D}_{C, K, S}\lp \x \rp = \operatorname{Concat}\lp \lbrace \mathcal{F}_i\lp \x_i \rp, \enskip \forall i \rbrace \rp,
    \end{gathered}
\end{equation}
where $\operatorname{Concat}(\cdot)$ performs the concatenation of all individual one-dimensional convolution outputs across the channel dimension. 
\end{definition}

\subsection{Encoder}
\label{sec:net_arch:encoder}
The encoder $\mathcal{E}$ architecture consists of a one-dimensional convolution with kernel size $\KE$ and stride equal to $\nicefrac{\KE}{2}$ similar to \cite{luo2019convTasNet}. Each convolved input audio-segment of $\KE$ samples is transformed to a $\CE$-dimensional vector representation where $\CE$ is the number of output channels of the $1D$-convolution. We force the output of the encoder to be strictly non-negative by applying a rectified linear unit (ReLU) activation on top of the output of the $1D$-convolution. Thus, the encoded input mixture representation could be expressed as:
\begin{equation}
\label{eq:encoder}
    \begin{gathered}
    \vx = \mathcal{E} \lp \textbf{x} \rp = \srelu \lp  \operatorname{Conv1D}_{\CE, \KE, \nicefrac{\KE}{2}} \lp \textbf{x}\rp  \rp  \in \R^{\CE \times L},
    \end{gathered}
\end{equation}
where the activation $\srelu \lp \cdot \rp $ is applied element-wise.

\subsection{Separator}
\label{sec:net_arch:separator}
In essence, the separator $\mathcal{S}$ module performs the following transformations to the encoded mixture representation $\vx \in \R^{\CE \times L}$:
\begin{enumerate}
    \item Projects the encoded mixture representation $\vx \in \R^{\CE \times L}$ to a new channel space through a layer-normalization (LN) \cite{ba2016layernormalization} followed by a point-wise convolution as shown next:
    \begin{equation}
    \label{eq:sep0}
    \begin{gathered}
    \y_0 = \operatorname{Conv1D}_{\Cout, 1, 1} \lp \operatorname{LN}\lp \vx \rp  \rp   \in \R^{\Cout \times L},
    \end{gathered}
    \end{equation}
    where $\operatorname{LN}\lp \vx \rp$ denotes a layer-normalization layer in which the moments used are extracted across the temporal dimension for each channel separately.
    \item Performs repetitive non-linear transformations provided by $B$ U-convolutional blocks (U-ConvBlocks) on the intermediate representation $\y_0$. In other words, the output of the $i$th U-ConvBlock would be denoted as $\y_i \in \R^{\Cout \times L}$ and would be used as input for the $(i+1)$th block. Each U-ConvBlock extracts and aggregates information from multiple resolutions which is extensively described in Section \ref{sec:net_arch:separator:main_block}.  
    \item Aggregates the information over multiple channels by applying a regular one-dimensional convolution for each source on the transposed feature representation $\y_B^\T \in \R^{L \times \Cout}$. Effectively, for the $i$th source we obtain an intermediate latent representation as shown next:
    \begin{equation}
    \label{eq:premasks}
    \begin{gathered}
    \mathbf{z}_i = \operatorname{Conv1D}_{\Cout, \CE, 1} \lp  \y_B^\T  \rp^\T   \in \R^{\CE \times L}
    \end{gathered}
    \end{equation}
    
    This step has been introduced in \cite{tzinis2019two} and empirically shown to make the training process more stable rather than using the activations from the final block $\y_B$ to estimate the masks.
    \item Combines the aforementioned latent codes for all sources $\mathbf{z}_i \enskip \forall i \in \{1, \cdots, N\}$ by performing a softmax operation in order to get mask estimates $\emi \in [0,1]^{\CE \times L}$ which add up to one across the dimension of the sources. Namely, the corresponding mask estimate for the $i$th source would be:
    \begin{equation}
    \label{eq:masks}
    \begin{gathered} 
    \emi = \operatorname{vec}^{-1} \lp \frac{\exp \lp \operatorname{ \operatorname{vec} \lp \mathbf{z}_i \rp} \rp } {\sum_{j=1}^{N} \exp \lp \operatorname{vec} \lp \mathbf{z}_j \rp \rp } \rp \in \R^{ \CE \times L},
    \end{gathered}
    \end{equation}
    where $\operatorname{vec} \lp \cdot \rp: \R^{K \times N} \rightarrow \R^{K \cdot N} $ and $\operatorname{vec}^{-1} \lp \cdot \rp: \R^{K \cdot N} \rightarrow \R^{K \times N} $ denotes the vectorization of an input tensor and the inverse operation, respectively. 
    \item Estimates a latent representation $\evi \in \R^{ \CE \times L}$ for each source by multiplying element-wise the encoded mixture representation $\vx$ with the corresponding mask $\emi$:
    \begin{equation}
    \label{eq:estimated_latent}
    \begin{gathered} 
    \evi = \vx \odot \emi \in \R^{ \CE \times L},
    \end{gathered}
    \end{equation}
    where $\mathbf{a} \odot \mathbf{b}$ is the element-wise multiplication of the two tensors $\mathbf{a} $ and $ \mathbf{b}$ assuming that they have the same shape. 
\end{enumerate}
%
\begin{figure}[!htb]
    \centering
      \includegraphics[width=\linewidth]{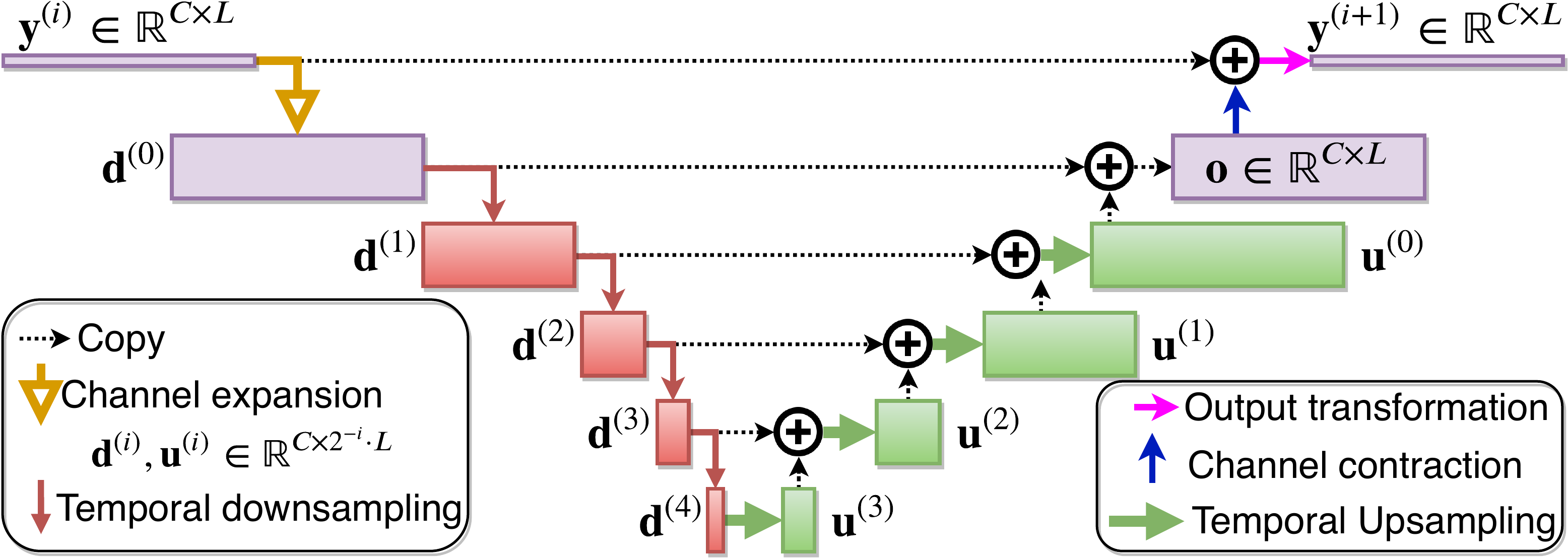}
      \caption{U-ConvBlock architecture.}
      \label{fig:uconvblock}
\end{figure}
\begin{algorithm}[t!]
\SetAlgoLined
\KwInput{$\y^{(i)} \in \R^{\Cout \times L}$} 
\KwOutput{$\y^{(i+1)} \in \R^{\Cout \times L}$} 
\tcp{Expand channel dimensions}
 $\mathbf{q} \gets \operatorname{PReLU}_{\Cin} \lp \operatorname{LN} \lp \operatorname{Conv1D}_{\Cin, 1, 1} \lp \y^{(i)} \rp \rp \rp $\;
 $\mathbf{d}^{(0)} \gets \operatorname{PReLU}_{\Cin} \lp \operatorname{LN} \lp \operatorname{DWConv1D}_{\Cin, \Kin, 1} \lp \mathbf{q} \rp \rp \rp $\;
 \For{$i = 1$; $i{+}{+}$; while $i <= Q$}{
 \tcp{Successive depth-wise downsampling}
  $\mathbf{d}^{(i)} \gets \operatorname{LN} \lp \operatorname{DWConv1D}_{\Cin, \Kin, \Sin} \lp \mathbf{d}^{(i-1)} \rp \rp $\;
  $\mathbf{d}^{(i)} \gets \operatorname{PReLU}_{\Cin} \lp \mathbf{d}^{(i)}  \rp $\;
 }
 $\mathbf{u}^{(Q)} \gets \mathbf{d}^{(Q)}$\;
 \For{$i = Q - 1$; $i{-}{-}$; while $i >= 0$}{
 \tcp{Upsample and add resolutions}
  $\mathbf{u}^{(i)} \gets \mathbf{d}^{(i)} + \mathcal{I}_{\Sin} \lp \mathbf{u}^{\lp i+1\rp} \rp $\;
 }
 $\mathbf{o} \gets  \operatorname{LN} \lp \operatorname{Conv1D}_{\Cout, 1, 1} \lp \operatorname{PReLU}_{\Cout} \lp \operatorname{LN} \lp \mathbf{u}^{(0)} \rp \rp \rp \rp $\;
 \Return $ \operatorname{PReLU}_{\Cout} \lp \y^{(i)} + \mathbf{o}  \rp$\;
 \caption{U-ConvBlock forward pass}
 \label{alg:uconvblock}
\end{algorithm}

\subsubsection{U-convolutional block (U-ConvBlock)}
\label{sec:net_arch:separator:main_block}
U-ConvBlock uses a block structure which resembles a depth-wise separable convolution \cite{sifre2014depthwiseseparable} with a skip connection as in ConvTasNet \cite{luo2019convTasNet}. However, instead of performing a regular depth-wise convolution as shown in \cite{chollet2017xception_depthwiseseparable} or a dilated depth-wise which has been successfully utilized for source separation \cite{luo2019convTasNet, tzinis2019improving, tzinis2019two} our proposed U-ConvBlock extracts information from multiple resolutions using $Q$ successive temporal downsampling and $Q$ upsampling operations similar to a U-Net architecture \cite{ronneberger2015original_unet}. More importantly the output of each block leaves the temporal resolution intact while increasing the effective receptive field of the network multiplicatively with each temporal sub-sampling operation \cite{luo2016effectivereceptivefield}. An abstract view of the $i$th U-ConvBlock is displayed in Figure \ref{fig:uconvblock} while a detailed description of the operations is presented in Algorithm \ref{alg:uconvblock}. 
\begin{definition}
$\operatorname{PReLU}_{C}:\R^{C \times L} \rightarrow \R^{C \times L}$ defines a parametric rectified linear unit (PReLU) \cite{he2015PReLU} with $C$ learnable parameters $\mathbf{a}  \in  \R^{C} $. When applied to an input matrix $\y \in \R^{C \times L}$ the non-linear transformation could be defined element-wise as:
\begin{equation}
\label{eq:prelu}
    \begin{gathered}
    \operatorname{PReLU}_{C}\lp \y \rp_{i,j} = \operatorname{max} \lp 0, \y_{i,j}  \rp + \mathbf{a}_i \cdot \operatorname{min} \lp 0, \y_{i,j}  \rp 
    \end{gathered}
\end{equation}
\end{definition}
\begin{definition}
$\mathcal{I}_{M}:\R^{C \times L} \rightarrow \R^{C \times M \cdot L}$ defines a nearest neighbor temporal interpolation by a factor of $M$. When applied on an input matrix $\y \in \R^{C \times L}$ this upsampling procedure could be formally expressed element-wise as: $\mathcal{I}_{M}\lp \mathbf{u} \rp_{i,j} = \mathbf{u}_{i, \left \lfloor{\nicefrac{j}{M}}\right \rceil}$
\end{definition}
\subsection{Decoder}
\label{sec:net_arch:decoder}
Our decoder module $\mathcal{D}$ is the final step in order to transform the latent space representation $\evi$ for each source back to the time-domain. In our proposed model we follow a similar approach as in \cite{tzinis2019two} where each latent source representation $\evi$ if fed through a different transposed convolution decoder $\operatorname{ConvTr1D}_{\CE, \KE, \nicefrac{\KE}{2}}$. The efficacy of dealing with different types of sources using multiple decoders has also been studied in \cite{differentdecoders}. Ignoring the permutation problem, for the $i$th source we have the following reconstruction in time: 
\begin{equation}
\label{eq:decoder}
    \begin{gathered}
    \esi = \mathcal{D}_i \left( \evi \right) = \operatorname{ConvTr1D}_{\CE, \KE, \nicefrac{\KE}{2}} \lp \evi \rp
    \end{gathered}
\end{equation}
\section{Experimental Setup}
\label{sec:exp_setup}
\subsection{Audio source separation tasks}
\label{sec:exp_setup:datasets}
\noindent\textbf{Speech separation:} We perform speech separation experiments in accordance with the publicly available WSJ0-2mix dataset \cite{hershey2016deepclustering} and other studies \cite{luo2019dual, zeghidour2020wavesplit, liu2019DeepCASA}. Speaker mixtures are generated by randomly mixing speech utterances with two active speakers at random signal to noise ratios (SNR)s between $-5$ and $5$dB from the Wall Street Journal (WSJ0) corpus \cite{WSJ0}.\\ 
\noindent\textbf{Non-speech sound separation:} For our non-speech sound separation experiments we follow the exact same setup as in \cite{tzinis2019two} and utilize audio clips from the environmental sound classification (ESC50) data collection \cite{esc50} which consists of a wide variety of sounds (non-speech human sounds, animal sounds, natural soundscapes, interior sounds and urban noises). For each data sample, two audio sources are mixed with a random SNR between $-2.5$ and $2.5$dB where each source belongs to a distinct sound category from a total of $50$. 
\subsection{Data preprocessing and generation}
We follow the same data augmentation process which was firstly introduced in \cite{tzinis2019two} and it has been show beneficial in other recent studies \cite{zeghidour2020wavesplit}. The process for generating a mixture is the following: A) random choosing two sound classes (for universal sound separation) or speakers (for speech separation) B) random cropping of $4$sec segments from two sources audio files C) mixing the source segments with a random SNR (as specified in Section \ref{sec:exp_setup:datasets}). For each epoch, $20,000$ new training mixtures are generated. Validation and test sets are generated once with each one containing $3,000$ mixtures. Moreover, we downsample each audio clip to $8$kHz, subtract its mean and divide with the standard deviation of the mixture. 
\subsection{Training and evaluation details}
\label{sec:exp_setup:train_eval}
All models are trained for $120$ epochs using a batch size equal to $4$. As a loss function we use the negative permutation-invariant \cite{Yu2017PIT} scale-invariant signal to distortion ratio (SI-SDR) \cite{le2019sdr} which is defined between the clean sources $\textbf{s}$ and the estimates $\widehat{\textbf{s}}$ as: 
\begin{equation}
\label{eq:SISDR}
    \begin{gathered}
    \mathcal{L} = - \text{SI-SDR}(\textbf{s}^*, \widehat{\textbf{s}}) = - 10 \log_{10} \left(\| \alpha \textbf{s}^*\|^2 / \| \alpha \textbf{s}^* - \widehat{\textbf{s}}\|^2 \right),
    \end{gathered}
\end{equation}
where $\textbf{s}^*$ denotes the permutation of the sources that maximizes SI-SDR and $\alpha =  \widehat{\textbf{s}}^\top  \textbf{s}^* /\|\textbf{s}\|^2$ is just a scalar. In order to evaluate the performance of our models we use the SI-SDR improvement (SI-SDRi) which is the gain that we get on SI-SDR measure using estimated signal instead of the mixture signal.
\subsection{\sudo configurations}
\label{sec:exp_setup:our_model_config}
For the encoder $\mathcal{E}$ and decoder modules $\mathcal{D}$ we use a kernel size $\KE=21$ corresponding to $2.625$ms and a number of basis equal to $\CE = 512$. For the configuration of each U-ConvBlock we set the input number of channels equal to $\Cout = 128$, the number of successive resampling operations equal to $Q = 4$ and the expanded number of channels equal to $\Cin = 512$. In each subsampling operation we reduce the temporal dimension by a factor of $2$ and all depth-wise separable convolutions have a kernel length of $\Kin=5$ and a stride of $\Sin=2$. We propose $3$ different models which are configured through the number $B$ of U-ConvBlocks inside the separator module $\mathcal{S}$. Namely, \sudol, \sudom, \sudos consist of $16$, $8$ and $4$ blocks, respectively. During training, we use the Adam optimizer \cite{adam} with an initial learning rate set to $0.001$ and we decrease it by a 
factor of $5$ every $50$ epochs.     
\subsection{Literature models configurations}
\label{sec:exp_setup:literature}
We compare against the best configurations of some of the latest state-of-the-art approaches for speech \cite{luo2019convTasNet, luo2019dual}, universal \cite{tzinis2019two} and music \cite{defossez2019demucs} source separation. For a fair comparison with the aforementioned models we use the authors original code, the best performing configurations for the proposed models as well as the suggested training process. For Demucs \cite{defossez2019demucs}, $80$ channels are used instead of $100$ in order to be able to train it on a single graphical processing unit (GPU).
\subsection{Measuring computational resources}
\label{sec:exp_setup:computational_resources}
One of the main goals of this study is to propose a model for audio source separation which could be trained using limited computational resources and deployed easily on a mobile or edge-computing device  \cite{lane2016deepx_deeplearning_onmobiledevs}. Specifically, we consider the following aspects which might cause a computational bottleneck during inference or training:
\begin{enumerate}
    \item Number of executed floating point operations (FLOPs).
     \item Number of trainable parameters.
    \item Memory allocation required on the device for a single pass.
    \item Time for completing each process. 
\end{enumerate}
We are using various sampling profilers in Python for tracing all the requirements on an Intel Xeon CPU E5-2695 v3 @ 2.30GHz CPU and a Nvidia Tesla K80 GPU.
\section{Results \& Discussion}
\label{sec:results}
\begin{figure}[!t]
    \centering
      \includegraphics[width=\linewidth]{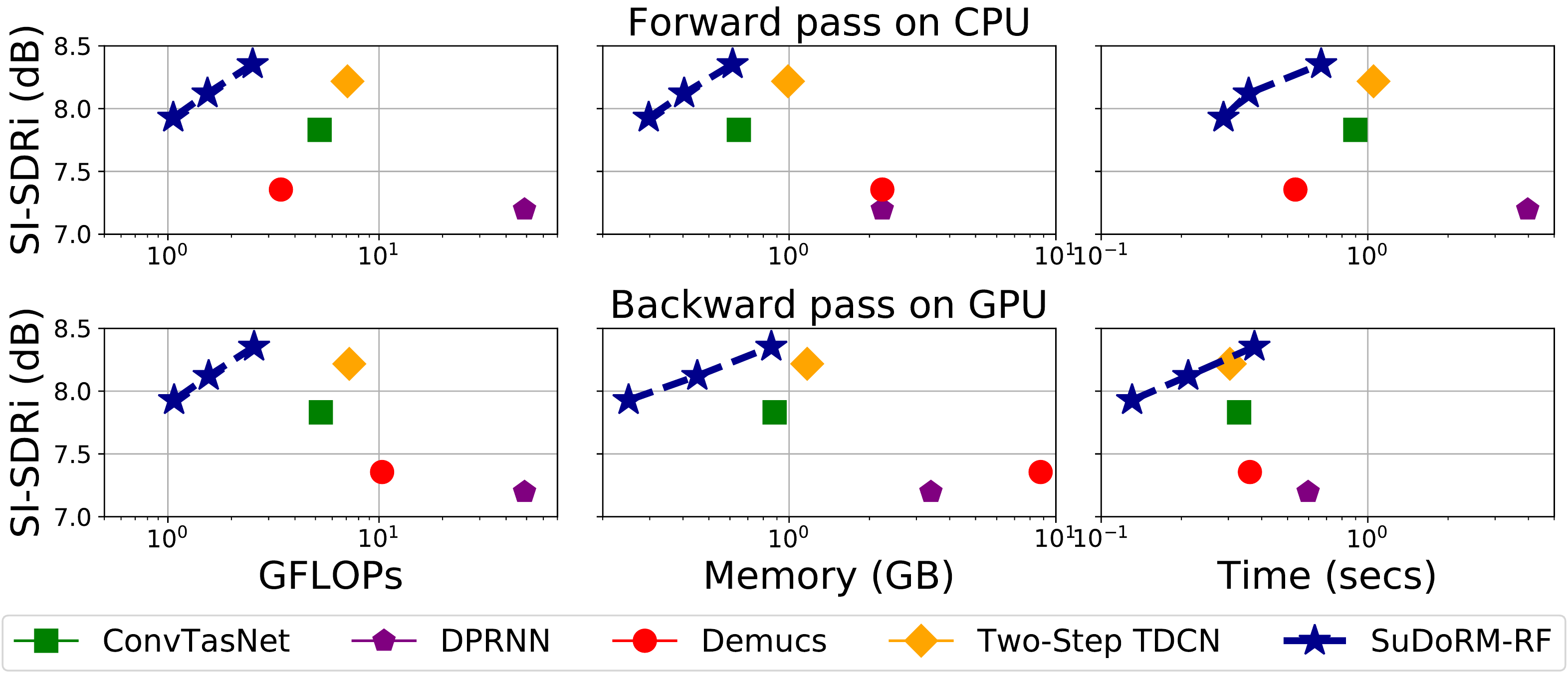}
      \caption{SI-SDRi non-speech sound separation performance on ESC50 vs computational resources with an input audio of $8000$ samples for all models. (Top row) computational requirements for a single forward pass on CPU (Bottom) for a backward pass on GPU. All x-axis are shown in log-scale while the $3$ connected blue stars correspond to the three \sudo configurations from Section \ref{sec:exp_setup:our_model_config}.}
      \label{fig:pareto}
    \vspace{-5pt}
\end{figure}

In Table \ref{tab:final_results}, we show the separation performance alongside computational requirements for some of the most recent state-of-the-art models in the literature and the proposed \sudo configurations. It is easy to see that the proposed models can match and even outperform the separation performance of other several state-of-the-art models using orders of magnitude less computational requirements. A better visualization for understanding the Pareto efficiency of the proposed architectures is displayed in Figure \ref{fig:pareto} where we show for each model its performance on non-speech sound separation vs a specific computational requirement. We do not show the same plots for speech separation on the WSJ dataset as the patterns were similar.  
\begin{table*}[!t]
    \centering
    \begin{tabular}{l|c|c|c|cc|cc|cc}
\toprule
\multirow{2}{*}{Model} & \multicolumn{2}{c|}{SI-SDRi (dB)} & Parameters & \multicolumn{2}{c|}{GFLOPs} & \multicolumn{2}{c|}{Memory (GB)} & \multicolumn{2}{c|}{Time (sec)} \\
 & Speech separation & Non-speech separation & (millions)& I & B & I & B & I & B\\
\hlinewd{1pt}
ConvTasNet \cite{luo2019convTasNet} & 15.30* & 7.74 & 5.05 & 5.23 & 5.30 & 0.65 & 0.88 & 0.90 & 0.33 \\
Demucs \cite{defossez2019demucs} & 12.12 & 7.23 & 415.09 & 3.43 & 10.34 & 2.24 & 8.77 & 0.53 & 0.36 \\
DPRNN \cite{luo2019dual} & 18.80* & 7.20 & 2.63 & 48.89 & 48.90 & 2.23 & 3.40 & 3.98 & 0.60 \\
Two-Step TDCN \cite{tzinis2019two} & 16.10* & 8.22 & 8.63 & 7.09 & 7.23 & 0.99 & 1.17 & 1.05 & 0.30 \\
\hlinewd{1pt}
\sudol & 17.02 & 8.35 & 2.66 & 2.52 & 2.56 & 0.61 & 0.86 & 0.67 & 0.38 \\
\sudom & 15.37 & 8.12 & 1.42 & 1.54 & 1.56 & 0.40 & 0.45 & 0.36 & 0.21 \\
\sudos & 13.39 & 7.93 & 0.79 & 1.06 & 1.07 & 0.30 & 0.25 & 0.29 & 0.13 \\
\bottomrule
\end{tabular}
\caption{SI-SDRi separation performance for all models on both separation tasks (speech and non-speech) alongside their computational requirements for performing inference on CPU (I) and a backward update step on GPU (B) for one second of input audio or equivalently $8000$ samples. * We assign the maximum SI-SDRi performance obtained by our runs and the reported number on the corresponding paper.}
\label{tab:final_results}
\end{table*}

\subsection{Floating point operations (FLOPs)}
Different devices (CPU, GPU, mobiles, etc.) have certain limitations on their FLOPs throughput capacity. In the case of an edge device, the computational resource one might be interested in is the number of FLOPs required during inference. On the other hand, training on cloud machines might be costly if a huge number of FLOPs is needed in order to achieve high separation performance. As a result, it is extremely important to be able to train and deploy models which require a low number of computations \cite{howard2017mobilenets}. We see from the first column of Figure \ref{fig:pareto} that \sudo models scale well as we increase the number of U-ConvBlocks $B$ from $4\rightarrow8\rightarrow16$. Furthermore, we see that for both forward and backward passes the family of the proposed \sudo models appear more Pareto efficient in terms of SI-SDRi performance vs Giga-FLOPs (GFLOPs) and time required compared to the other state-of-the-art models which we take into account. Specifically, the DPRNN model \cite{luo2019dual} which performs sequential matrix multiplications (even with a low number of parameters) requires at least $45$ times more FLOPs for a single pass compared to \sudos while performing worse when trained for the same number of epochs.
\subsubsection{Cost-efficient training}
\begin{figure}[!htb]
    \centering
      \includegraphics[width=\linewidth]{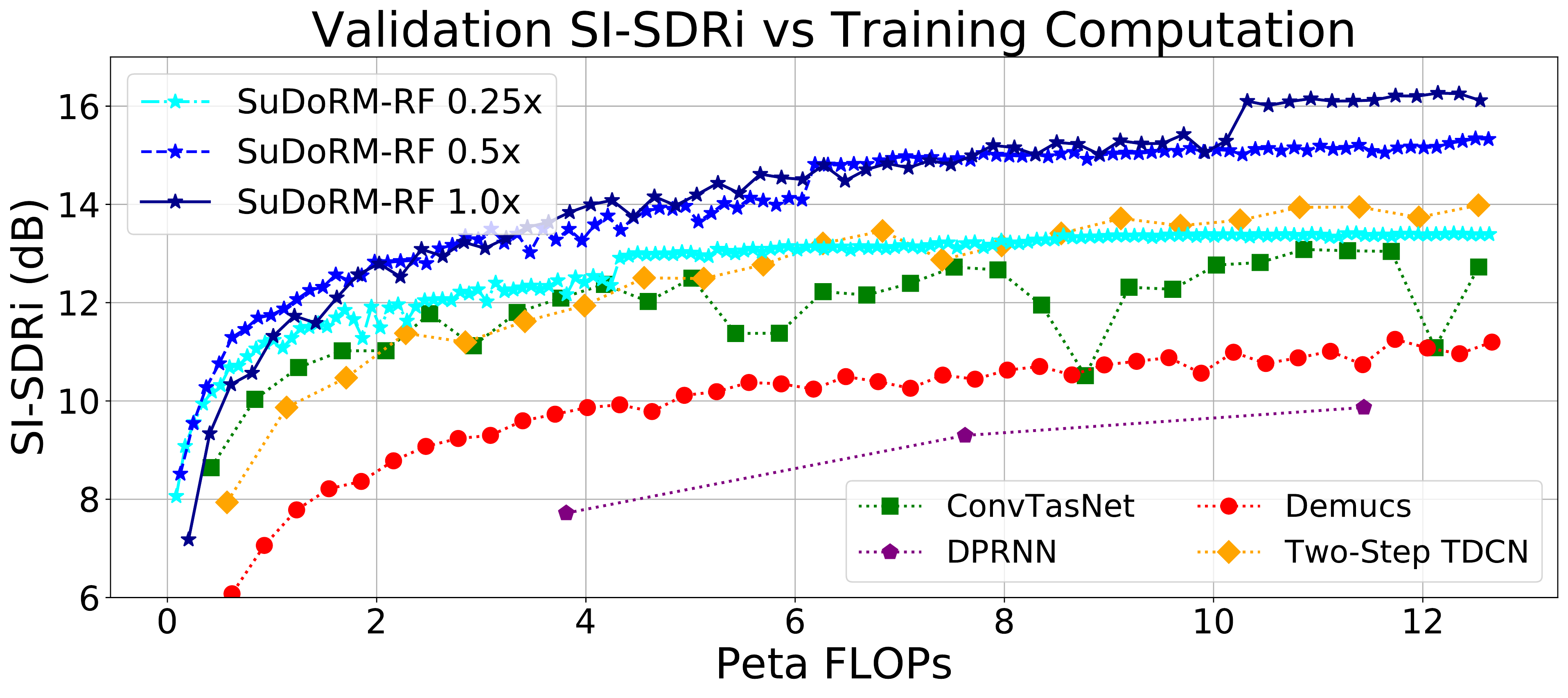}
      \caption{Validation SI-SDRi separation performance for speech-separation vs the number of FLOPs executed during training. All models are trained using batches of $4$ mixtures with $32,000$ time-samples each. Each point corresponds to a completed training epoch.}
      \label{fig:costefficienttraining}
    \vspace{-10pt}
\end{figure}

Usually one of the most detrimental factors for training deep learning models is the requirement of allocating multiple GPU devices for several days or weeks until an adequate performance is obtained on the validation set. This huge power consumption could lead to huge cloud services rental costs and carbon dioxide emissions \cite{cai2019onceandforall}. In Figure \ref{fig:costefficienttraining}, we show the validation SI-SDRi performance for the speech separation task which is obtained by each model versus the total amount of FLOPs performed. For each training epoch all models perform updates while iterating over $20,000$ audio mixtures. Notably, \sudo models outperform all other models in terms of cost-efficient training as they obtain better separation performance while requiring significantly less amount of training FLOPs. For instance, \sudol obtains $\approx 16$dB in terms of SI-SDRi compared to $\approx 10$dB of DPRNN \cite{luo2019dual} which manages to complete only $3$ epochs given the same number of training FLOPs.   

\subsection{Trainable parameters}
From Table \ref{tab:final_results} it is easy to see that \sudo architectures are using orders of magnitude fewer parameters compared to the U-net architectures like Demucs \cite{defossez2019demucs} where each temporal downsampling is followed by a proportional increase to the number of channels. Moreover, the upsampling procedure inside each U-ConvBlock does not require any additional parameters. The \sudo models seem to increase their effective receptive field with significantly fewer parameters compared to dilated convolutional architectures \cite{luo2019convTasNet, tzinis2019two}.  Notably, our largest model \sudol matches the relatively low number of parameters of the DPRNN \cite{luo2019dual} model which is based on stacked RNN layers.

\subsection{Memory requirements}
In most of the studies where efficient architectures are introduced \cite{howard2017mobilenets, chollet2017xception_depthwiseseparable, mehta2019espnetv2, yu2019slimmablenets} authors are mainly concerned with the total number of trainable parameters of the network. The same applies to efficient architectures for source separation \cite{luo2019convTasNet, luo2019dual, maldonado2020lightweight}. However, the trainable parameters is only a small portion of total amount of memory required for a single forward or backward pass. The space complexity could easily blow up by the storage of intermediate representations. The latter could become even worse when multiple skip connections are present, gradients from multiple layers have to be stored or implementations require augmented matrices (dilated, transposed convolutions, etc.). In Figure \ref{fig:pareto}, we see that \sudo models are more pareto-efficient in terms of the memory required compared to the dilated convolutional architectures of ConvTasNet \cite{luo2019convTasNet} and Two-Step TDCN \cite{tzinis2019two} where they require an increased network depth in order to increase their receptive field. Although \sudo models do not perform downsampling in every feature extraction step as Demucs \cite{defossez2019demucs} does, we see that the proposed models require orders of magnitude less memory especially during a backward update step as the number of parameters in Demucs is significantly higher. Finally, \sudo models have a smaller memory footprint because the encoder $\mathcal{E}$ performs a temporal downsampling by a factor of $\operatorname{div}\lp \KE, 2\rp=10$ compared to DPRNN \cite{luo2019dual} which does not reduce the temporal resolution at all.

\subsection{Ablation study on WSJ0-2mix}
We perform a small ablation study in order to show how different parameter choices in \sudo models affect the separation performance. In order to be directly comparable with the numbers reported by several other studies  \cite{luo2019convTasNet, luo2019dual, zeghidour2020wavesplit, liu2019DeepCASA}, we train our models for $200$ epochs and test them using the given data splits from WSJ0-2mix dataset \cite{hershey2016deepclustering}. The results are shown in Table \ref{tab:ablation_study}.

\begin{table}[!t]
    \centering
    \begin{tabular}{c|c|c|c|c|c|c|c}
\toprule
$\KE$ & $\Cout$ & $B$ & $Q$ & Norm & Mask Act. & Dec. & SI-SDRi \\
\hlinewd{1pt}
21 & 128 & 16 & 4 & LN & Softmax & 2 & 16.0 \\
\hline
17 & 128 & 16 & 4& LN & ReLU & 1 & 15.9 \\
\hline
17 & 128 & 16 & 4& GLN & ReLU & 1 & 16.8 \\
\hline
21 & 256 & 20 & 4& GLN & ReLU & 1 & 17.7 \\
\hline
41 & 256 & 32 & 4 & GLN & ReLU & 1 & 17.1 \\
\hline
41 & 256 & 20 & 4 & GLN & ReLU & 1 & 16.8 \\
\hline
21 & 512 & 18 & 7 & GLN & ReLU & 1 & 18.0 \\
\hline
21 & 512 & 20 & 2 & GLN & ReLU & 1 & 17.4 \\
\hline
21 & 512 & 34 & 4 & GLN & ReLU & 1 & 18.9 \\
\bottomrule
\end{tabular}
\caption{SI-SDRi separation performance on WSJ0-2mix for various parameter configurations of \sudo models. Mask Act. corresponds to the activation function before the mask estimation and Dec. specifies the number of decoders we are using before reconstructing the time-domain signals. GLN corresponds to the global layer normalization as described in \cite{luo2019convTasNet}. All the other parameters have the same values as described in Section \ref{sec:exp_setup:our_model_config}}
\label{tab:ablation_study}
 \vspace{-10pt}
\end{table}

\section{Conclusions}
\label{sec:conclusions}
In this study, we have introduced the \sudo network, a novel architecture for efficient universal sound source separation. The proposed model is capable of extracting multi-resolution temporal features through successive depth-wise convolutional downsampling of intermediate representations and aggregates them using a non-parametric interpolation scheme. In this way, \sudo models are able to significantly reduce the required number of layers in order to effectively capture long-term temporal dependencies. We show that these models can perform similarly or even better than recent state-of-the-art models while requiring significantly less computational resources in FLOPs, memory and time. In the future, we aim to use \sudo models for real-time low-cost source separation.

\bibliographystyle{IEEEbib}
\bibliography{refs}

\end{document}